\documentstyle[11pt]{article}

\newcommand{\order}{{\cal O}}
\newcommand{\be}{\begin{equation}}
\newcommand{\ee}{\end{equation}}
\newcommand{\bea}{\begin{eqnarray}}
\newcommand{\eea}{\end{eqnarray}}
\newcommand{\nl}{\\ \nonumber}

\def\lesssim{\mathrel{\raise.3ex\hbox{$<$\kern-.75em\lower1ex\hbox{$\sim$}}}}
\def\gtrsim{\mathrel{\raise.3ex\hbox{$>$\kern-.75em\lower1ex\hbox{$\sim$}}}}

\begin{document}
\title{Nonperturbative Techniques for QED Bound States
\thanks{Talk presented at the MRST conference, University of Rochester, 8-9 May, 2000.}}

\author{Richard Hill\thanks{electronic address: rjh@mail.lns.cornell.edu}\\
\small Newman Laboratory of Nuclear Studies, Cornell University\\
\small Ithaca, NY 14853.
}

\maketitle

\begin{abstract}
Advantages of using a low-energy effective theory to study bound
state properties are briefly discussed, and a nonperturbative implementation
of such an effective theory is described within the context of nonrelativistic 
quantum mechanics.  The hydrogen atom, in the approximation of a 
structureless, infinite-mass nucleus, but with the leading
relativistic and radiative corrections included, is used to demonstrate the 
construction and solution of the effective theory. The resulting 
Hamiltonian incorporates a finite ultraviolet cutoff and can be solved
nonperturbatively.  
An appendix lists explicit formulae for the various matrix
elements necessary to diagonalize the Hamiltonian using gaussian basis sets. 
\end{abstract}

\section{Introduction}
The study of QED bound states tests our understanding of nonperturbative
field theory, and of the techniques used to study these nonperturbative
phenomena.  Comparison of theoretical predictions to precision 
experiments gives the best determination of various fundamental
constants (e.g.  $R_\infty$ from hydrogen spectral lines, $m_\mu$
from the muonium hyperfine splitting).  Once all of the relevant
constants are determined, a disagreement in 
the theory-experiment comparison can act 
as a signal to new physics.  

Given these motivations to study precision bound state QED, there
are some difficulties which must be addressed.  Two of these
difficulties are the complexity of relativistic formalisms, 
and the cumbersome nature of bound state perturbation theory. 

Traditionally
(although this tradition appears to be shifting), precision 
bound state calculations have been performed in relativistic 
formalisms, using the Bethe-Salpeter equation or one of its
variants.  A drawback of the integrated approach of such a 
formalism, 
which does not distinguish between low-momentum and high-momentum 
contributions (since this would violate relativistic invariance),
is the difficulty in choosing an optimal gauge.  For instance,
Coulomb gauge could be most effective for parts of the calculation
involving low momentum, but is cumbersome for high-momentum where
Feynman gauge may be more appropriate.  However in Feynman gauge, 
spurious low-order contributions appear which,
though cancelling in the end, make it difficult to determine
exactly which Feynman diagrams must be evaluated to obtain a given
accuracy.  
When a low energy effective theory is used to separate high- and
low-momentum modes in a systematic way, the appropriate gauge
can be chosen independently for both momentum regions.  
Technical simplifications of this sort result
from the effective theory taking advantage of the essentially 
nonrelativistic character of the system.

Bound state perturbation theory involves nontrivial sums
over intermediate states, and in high orders 
the terms to be evaluated increase in number and complexity. 
Another problem is the reliance on an unperturbed state
which forms the basis for perturbations.  This is not
such a problem in the single particle case, where the familiar
Schr\"{o}dinger-Coulomb wavefunctions are an obvious choice,
but for multiparticle systems such as helium no analytic 
unperturbed wavefunction is known.  These difficulties can
be overcome by solving the effective theory 
nonperturbatively~\footnote{ The low-energy effective theory will still be
determined perturbatively, as an expansion in $\alpha$ 
and the typical atomic velocity $v$.  It is the 
effective theory, and not QED itself, which will then
be solved nonperturbatively.  }.

In the rest of the talk, after mentioning some examples
of low-energy effective theories, I 
demonstrate, in stages, the construction of an effective
Hamiltonian for the hydrogen atom with leading relativistic
and radiative corrections, in the approximation of 
a structureless, infinite mass nucleus\footnote{
The ideas here are in most cases a rewording, and
in others a slight extension of the presentation 
in Ref.\cite{Renormalizing}.   
}.   
Ref.~\cite{ops} gives a more modern application, 
to the decay rate of orthopositronium.  

\section{Low-Energy Effective Theories}

An example of a low-energy effective theory is NRQED 
(nonrelativistic QED)
field theory, described by the Lagrangian
\bea
\label{eq: NRQED}
{\cal L}_{\rm eff} &=& -\frac{1}{2}(E^2-B^2) 
	+ \psi^\dagger \left( i\partial_t -e\phi +\frac{D^2}{2m} \right.\nl
&&\left.+c_1\frac{D^4}{8m^3} 
	+c_2 \frac{e}{2m} \sigma\cdot B 
	+c_3 \frac{e}{8m^2} \nabla\cdot E
	+c_4 \frac{e}{8m^2}\{iD\cdot E\times\sigma\}+\cdots\right) \psi \nl
&&+(\psi\leftrightarrow\chi)\nl
&&+\frac{d_1}{mM}\psi^\dagger\sigma\psi\cdot\chi^\dagger\sigma\chi
	+\frac{d_2}{mM}\psi^\dagger\psi\chi^\dagger\chi +\dots ,
\eea
together with a cutoff prescription, at momentum scale $\Lambda$. 
Here $D=\nabla+ieA$ is the covariant derivative and $\psi,\chi$ are Pauli
spinor fields. $c_1,c_2,\dots,d_1,d_2,\dots$ are renormalization constants
which must be determined by matching the predictions of NRQED to 
those of QED.  The local operators 
parameterized by $d_1,d_2,\dots$ account for the short-distance/
high-momentum states which are excluded by the cutoff~\cite{NRQED}.

Another, and familiar, example of a low-energy effective
theory is nonrelativistic quantum mechanics, described by a
Hamiltonian,
\bea
\label{eq: Heff}
H_{\rm eff} &=& \frac{p^2}{2m} + V_{\rm long-range}(r) 
	+ \frac{d_1}{m^2} \delta_\Lambda^3(r) 
	+ \frac{d_2}{m^4} (-\nabla^2\delta_\Lambda^3(r))
	+ \frac{d_3}{m^4} {\bf p}\cdot \delta_\Lambda^3(r){\bf p}
	+\dots .
\eea
$V_{\rm long-range}(r)$ includes all long range interactions 
(the analogue of those operators parameterized by $c_i$ in Eq.(\ref{eq: NRQED}) ).
The remaining operators are the leading terms in an expansion  
of local operators.  The structure of $V_{\rm long-range}(r)$ and
the values of the coefficients $d_i$ are determined by matching 
predictions of the effective theory to those of the true theory 
(in the present case, QED)~ 
\footnote{ 
Alternatively, NRQED field theory
could be used in an intermediate step:  NRQED matched to QED,
QM matched to NRQED.
}.
It is on the effective Hamiltonian which we will focus, and for
which we will formulate our nonperturbative implementation.  

\section{The Schr\"{o}dinger-Coulomb Problem}
First, consider the Schr\"{o}dinger-Coulomb problem, where the
``true'' theory has Hamiltonian 
\be
H_{\rm SC} = \frac{p^2}{2m} -\frac{(Z\alpha)}{r}.
\ee
We will treat this as an example problem, to demonstrate the
basic ideas in a familiar setting.  

Our goal here is to build an effective Hamiltonian in the 
form of Eq.(\ref{eq: Heff}), which describes the theory 
defined by $H_{\rm SC}$.  We first introduce a 
cutoff $\Lambda\approx m$ which effectively removes 
states of momentum $p\gtrsim\Lambda$ from the theory.  
This is reasonable, since for such large momenta, 
$e^+e^-$ pair creation and other relativistic effects 
become important in QED~ 
\footnote{
It happens that $H_{\rm SC}$ is
well defined without a UV cutoff, so that introducing
one may seem artificial here.  However, when
relativistic corrections are included, a cutoff is 
essential.  And even in the present case, there are 
numerical advantages to having the cutoff in place\,---\,for example, fewer 
basis functions are required in the matrix
diagonalization.
}.
The particular cutoff we choose is in the form of a gaussian
multiplying the momentum space potential:
\bea
\frac{1}{r}&\to& \left(\frac{1}{r}\right)_\Lambda \nl
\frac{4\pi}{q^2}&\to& \frac{4\pi}{q^2}e^{-\frac{q^2}{2\Lambda^2}},
\eea
where the second line is the Fourier transform of the first:
\be
\left(\frac{1}{r}\right)_\Lambda 
= \int\frac{d^3q}{(2\pi)^3}e^{i{\bf q}\cdot {\bf r}}\frac{4\pi}{q^2}
	e^{-\frac{q^2}{2\Lambda^2}} 
= \frac{1}{r}{\rm erf}(\frac{\Lambda r}{\sqrt{2}}),
\ee
with ${\rm erf}(x)$ the error function.  Introducing 
the cutoff modifies the Coulomb potential at small distances
$r\lesssim 1/\Lambda$.  With this form of the cutoff, a 
natural choice for the local operator $\delta_\Lambda^3(r)$ is:
\be
\delta_\Lambda^3(r) 
= \int\frac{d^3q}{(2\pi)^3}e^{i{\bf q}\cdot {\bf r}}
	e^{-\frac{q^2}{2\Lambda^2}} 
= \frac{\Lambda^3}{(2\pi)^{3/2}}e^{-\frac{\Lambda^2 r^2}{2}}
\ee
The other local operators in Eq.(\ref{eq: Heff}) are derived
from $\delta_\Lambda^3(r)$ by differentiation.  

With our modified Hamiltonian, we no longer expect to solve
the eigenvalue problem for the energies analytically; instead,
we look for a numerical solution~\footnote
{Since an analytic solution takes the form of a series 
expansion in $\alpha$, a numerical approach is more natural
as a nonperturbative tool. 
}.   
Two approaches can be implemented easily:
\newline(i) {\bf Differential Equation Integration}:  With $H_{\rm eff}$ expressed
as a differential operator, the Schr\"{o}dinger
equation is:
\be
\left(-\frac{1}{2m}\nabla^2 + V(r) \right)\psi(r) = E\psi(r) .
\ee
Together with appropriate boundary conditions ($\psi(r\to\infty)=0$,
$\psi(r\to 0)$ finite), this eigenvalue problem can be solved
using standard numerical differential equation integration routines. 

\noindent(ii) {\bf Matrix Diagonalization}:  This technique will 
be our main focus, 
it being more readily adaptable than the differential equation method
when radiative corrections are introduced. 
Elements of a (linearly independent, but not necessarily orthogonal) 
basis are first chosen, 
for example:
\be
\Phi^i_{lm}({\bf r}) = Y_{lm}(\theta,\phi) r^l e^{-\frac{r^2}{2R_i^2}}.
\ee
where $Y_{lm}$ is a spherical harmonic and 
$R_i$ runs over an appropriate range of distance scales. 
We now take matrix elements of all operators involved:
(for clarity, we replace the three indices $i,l,m$ above by a single index)
\bea
\langle \Phi^i |p^2|\Phi^j\rangle &\equiv& (p^2)^{ij} \nl
\langle \Phi^i | V |\Phi^j\rangle &\equiv&  V^{ij} \nl
\langle \Phi^i|\Phi^j\rangle &\equiv& W^{ij}.
\eea
Now taking matrix elements of the Schr\"{o}dinger equation, we obtain:
\be
\left( \frac{1}{2m}(p^2)^{ij} + V^{ij} \right) \psi_j = E W^{ij}\psi_j,
\ee
or as matrices:
\be 
\label{eq: matrix}
\left( \frac{1}{2m}p^2 + V \right)\psi = E W \psi,
\ee
where $\Psi = \Phi^j \psi_j$, and repeated indices are summed over. 
Since we have not taken our basis elements to be orthonormal, 
$W^{ij} \neq \delta^{ij}$.  To recover a more standard form, 
define, as matrices, 
\bea
W^{-\frac{1}{2}} p^2 W^{-\frac{1}{2}} &\equiv& \tilde{p^2}\nl
W^{-\frac{1}{2}} V W^{-\frac{1}{2}} &\equiv& \tilde{V}\nl
W^{\frac{1}{2}} \psi &\equiv& \tilde{\psi}.
\eea
Then Eq.(\ref{eq: matrix}), upon multiplying both sides by $W^{-\frac{1}{2}}$
becomes:
\be
\left( \frac{1}{2m}\tilde{p^2} + \tilde{V} \right)\tilde{\psi} 
= E \tilde{\psi}.
\ee
This matrix eigenvalue equation can be solved easily using 
standard matrix diagonalization routines\footnote{
See for example Ref.~\cite{NRecipes}.  For the relatively small
computer time required for the problems treated here, a commercial
program like ``Maple'' or ``Mathematica'' works fine.  
}.  

The long range potential of Eq.(\ref{eq: Heff}) is 
\be
V_{\rm long-range}(r) = -(Z\alpha)\left(\frac{1}{r}\right)_\Lambda
	\equiv V_C.
\ee
To complete the determination of $H_{\rm eff}$, we must 
evaluate the necessary local operator coefficients $d_1,d_2,\dots$. 
First, we should decide which coefficients must be evaluated, 
and to what accuracy.  This is accomplished by noting that 
\be
\langle \psi_0|\frac{d_1}{m^2}\delta_\Lambda^3(r)|\psi_0\rangle
\approx \frac{d_1}{m^2}|\psi_0|^2 \approx d_1 m\alpha^3 .
\ee
So, to determine energy levels through, say, $\order(m\alpha^6)$,
we need determine $d_1$ through $\order(\alpha^3)$.  Terms
parameterized by $d_2$ and $d_3$ contain two more powers of 
$p/m\approx\alpha$, and so contribute at $\order(d_2m\alpha^5)$,
$\order(d_3m\alpha^5)$ to the energy.  

We will implement the matching procedure between the true and
effective theory by perturbative matching of scattering 
amplitudes for the process ${\bf k}\to {\bf l}$.  
The lowest order scattering amplitude (the Born term)
is simply the momentum space potential: (here ${\bf q}={\bf l}-{\bf k}$ 
is the momentum transfer)
\bea
\langle {\bf l}|T|{\bf k}\rangle 
&=& \left(-\frac{4\pi(Z\alpha)}{q^2}
	+\frac{d_1}{m^2} 
	+ \frac{d_2}{m^4}q^2
	+ \frac{d_3}{m^4}{\bf l}\cdot {\bf k}\right)
		e^{-\frac{q^2}{2\Lambda^2}} \nl
&=& -\frac{4\pi(Z\alpha)}{q^2}
	+\left(\frac{d_1}{m^2} +\frac{2\pi(Z\alpha)}{\Lambda^2}\right)
	+\left(\frac{d_2}{m^4} -\frac{\pi(Z\alpha)}{2\Lambda^4}\right)q^2
	+\frac{d_3}{m^4}{\bf l}\cdot {\bf k} + \order(k^4).
\eea
The first term matches the $\order(\alpha)$ result of the ``true'' 
Coulomb theory.  Requiring the other terms to vanish, through $\order(\alpha)$,
gives:
\bea
\label{eq: d11}
d_1^{(1)} &=& -2\pi\frac{m^2}{\Lambda^2} \\
d_2^{(1)} &=& -\frac{\pi}{2}\frac{m^4}{\Lambda^4} \\
d_3^{(1)} &=& 0,
\eea
where $d_1=(Z\alpha) d_1^{(1)} + (Z\alpha)^2 d_2^{(2)} +(Z\alpha)^3 d_3^{(3)} +\dots$,
$d_2=(Z\alpha) d_2^{(1)} +\dots$, $d_3=(Z\alpha) d_3^{(1)}+\dots$.   

At $\order(\alpha^2)$, our power counting shows that we need only 
evaluate $d_1^{(2)}$.  To do so, we consider threshold scattering
amplitudes in the true and effective theories~
\footnote{
Here we are using the relation $T_{lk}=V_{lk}+V_{lp}G_p(E)T_{pk}$, where
$G_p(E)=(E-p^2/(2m)+i\epsilon)^{-1}$ is the nonrelativistic 
propogator and for the scattering state, $E=k^2/(2m)$.
}:

\bea
\label{eq: T2}
&&\lim_{k\to 0} \langle {\bf l}|T_{\rm true}^{(2)}|{\bf k}\rangle 
	- \langle {\bf l}|T_{\rm eff}^{(2)}|{\bf k}\rangle
= \int\frac{d^3p}{(2\pi)^3}\left[\frac{-4\pi}{p^2}\frac{-2m}{p^2}
	\frac{-4\pi}{p^2} \right.\nl
&&\left.	
	-\left(-\frac{4\pi}{p^2}+\frac{d_1^{(1)}}{m^2}
		+\frac{d_2^{(1)}}{m^4}p^2\right)
	e^{-\frac{p^2}{2\Lambda^2}}\left(\frac{-2m}{p^2}\right)
	\left(-\frac{4\pi}{p^2}+\frac{d_1^{(1)}}{m^2}
		+\frac{d_2^{(1)}}{m^4}p^2\right)
	e^{-\frac{p^2}{2\Lambda^2}}
\right] +\frac{d_1^{(2)}}{m^2} 
\eea
Requiring this quantity to vanish yields:
\bea
\label{eq: d12}
d_1^{(2)} &=&\sqrt{\pi}\left(-\frac{10}{3}\left(\frac{m}{\Lambda}\right)^3
	-20\frac{d_2^{(1)}}{4\pi}\frac{\Lambda}{m}
	+6\left(\frac{d_2^{(1)}}{4\pi}\right)^2\left(\frac{\Lambda}{m}\right)^5
	\right) \nl
&=& -\frac{71\sqrt{\pi}}{96}\left(\frac{m}{\Lambda}\right)^3.
\eea
By a similar calculation at $\order(\alpha^3)$, with the above
calculated values of $d_2^{(1)}$ and $d_1^{(2)}$~
\footnote{
For energy levels correct through $\order(m\alpha^5)$,
it would be sufficient to include only $d_1^{(1)}$ and $d_1^{(2)}$,
with $d_2=0$. The value of $d_1^{(2)}$ would then be 
given by Eq.(\ref{eq: d12}) at $d_2^{(1)}=0$.
},
\be
d_1^{(3)} = -\left(\frac{5\sqrt{3}}{72}+\frac{1633}{384}\right)
	\left(\frac{m}{\Lambda}\right)^4.
\ee
\begin{table}
\begin{center}
\begin{tabular}{lll}
\hline
$n$ & $E/E_0$ & $E/E_0$ \\
\hline
$1$ & $0.99850$ & $1.0000000035$ \\
$2$ & $0.24981$ & $0.2499999996$ \\
$3$ & $0.11106$ & $0.1111111109$ \\
\hline
\end{tabular}
\caption{$S$~-state energy levels at $\alpha=0.02$,$\Lambda=m$.  
Here $E_0=-m\alpha^2/2$ is the ground state Coulomb energy.  
In the first column, $d_1=d_2=0$. In the second, 
these parameters take the values calculated in the text. 
}
\label{table: coulomb}
\end{center}
\end{table}
A comparison of the first few $S$~-state energy levels calculated
both with and without counterterms are shown in 
Table~\ref{table: coulomb}.  As expected, with counterterms included
the energy levels are in agreement with the Coulomb spectrum through
$\order(m\alpha^6)$.  This matching procedure can be extended
systematically to higher orders.

\section{Relativistic Corrections}

The leading relativistic corrections to the Schr\"{o}dinger-Coulomb
Hamiltonian take the form~\cite{Bethe}:
\be
\label{eq:Breit}
\delta H = -\frac{p^4}{8m^3} +\frac{\pi(Z\alpha)}{2m^2}\delta^3(r)
	+\frac{(Z\alpha)}{4m^2}\frac{{\bf L}\cdot{\bf \sigma}}{r^3}.
\ee

We can immediately see problems with these corrections as they
stand, if we are to solve nonperturbatively.   At high
momentum, the $p^4$ term dominates over $p^2$, causing the 
Hamiltonian spectrum to be unbounded from below.  Also, 
the $\delta$-function is too singular and all second-and
higher order perturbations involving this term will be
divergent.  To remedy these problems, we introduce a 
cutoff:
\bea
\delta^3(r)&\to& \delta^3_\Lambda(r), \\
\frac{-1}{r^3}=\frac{1}{r}\left(\frac{1}{r}\right)^\prime&\to&\frac{1}{r}
	\left(\frac{1}{r}\right)_\Lambda^\prime,
\eea
giving the cutoff Darwin and spin-orbit potentials:
\bea
\label{eq: Darwin}
V_D = \frac{\pi(Z\alpha)}{2m^2}\delta_\Lambda^3(r)\\
\label{eq: spinorbit}
V_{SO} = -\frac{(Z\alpha)}{4m^2}\frac{1}{r}
		\left(\frac{1}{r}\right)_\Lambda^\prime 
			{\bf L}\cdot{\bf \sigma}.
\eea
For the $p^4$ term, we choose to work with an equivalent form~
\cite{Renormalizing}~
\footnote{Here we could also apply the cutoff directly to the $p^4$ operator,
for example $p^4{\rm exp}(-p^2/\Lambda^2)$.  
The other form 
happens to make evaluation of the counterterms more convenient
since then only the free propogator $p^2/(2m)$ appears in scattering 
calculations.   
}:
\be
p^4\to (2m)^2(E-V)^2,
\ee
which defines our kinetic energy correction potential:
\be
\label{eq: kinetic}
V_K(E) = -\frac{1}{2m}(E-V_C)^2,
\ee
where $V_C$ is the cutoff Coulomb potential.   

Working through $\order(m\alpha^5)$, the only counterterms necessary
are $d_1^{(1)}$, which is given in Eq.(\ref{eq: d11}),
and $d_1^{(2)}$, which must be recalculated with the relativistic 
corrections in place:
\be
\label{eq: d12rel}
d_1^{(2)} = \sqrt{\pi}\left(-\frac{10}{3}\left(\frac{m}{\Lambda}\right)^3
-\frac{m}{\Lambda} +\frac{1}{8}\frac{\Lambda}{m}\right).
\ee
(This calculation is similar to that of
Eq.(\ref{eq: T2}), but with $T_{\rm true}$ now referring to the
scattering amplitude of QED in the external field approximation.)  
 
\section{Radiative Corrections}

It remains to include the effects of radiative corrections---in 
the language of Coulomb gauge, the effect of the 
transverse radiation field.  Our analysis is complicated here
by the fact that soft photons, of energy $E\sim m\alpha^2$,
can no longer be described by instantaneous potentials in
the electron's Hamiltonian, since their characteristic
propogation time ($\delta t\approx 1/E$) 
can be comparable to bound state timescales ($\delta t\approx 1/(m\alpha^2)$). 
We can deal with this by expanding the
state space to include two channels:  one with just the
electron, another with the electron and a transverse photon.  
The wavefunction now has components in both sectors: 
\be
\psi =
\left(
\begin{array}{c}
\psi_e \\
\psi_{e\gamma}
\end{array}
\right).
\ee

The new terms in the Hamiltonian which describe the coupling
between channels are:
\be
\delta H = H_\gamma -\frac{e}{m}{\bf p}\cdot {\bf A}({\bf r}),
\ee
where
\be
\label{eq: Hg}
H_\gamma = \int\frac{d^3q}{(2\pi)^3}
		\sum_{{\bf \epsilon}({\bf q})}\omega_{q}\,
	a_{{\bf q},{\bf \epsilon}({\bf q})}^\dagger\,	
		a_{{\bf q},{\bf \epsilon}({\bf q})}
\ee
is the photon kinetic energy operator, and
\be
{\bf A}({\bf r}) = \int\frac{d^3q}{(2\pi)^3}\sum_{{\bf \epsilon}({\bf q})}
		\frac{1}{\sqrt{2\omega_q}}
	\left(a_{q,\epsilon(q)} e^{i{\bf q}\cdot {\bf r}} + h.c \right) .  
\ee
For photons of momentum $q\ll m\alpha$ (i.e. $q\lesssim m\alpha^2$),
the multipole expansion is valid: 
\be 
{\bf A}({\bf r}) = {\bf A}(0) + {\bf r}\cdot{\bf \nabla} {\bf A}(0) +\dots
\ee
For $q\gg m\alpha^2$ (i.e. $q\gtrsim m\alpha$), the photon
propogation is effectively instantaneous, and can be described 
by instantaneous potentials in the electron Hamiltonian.  So,
denoting the potentials already present (Coulomb and relativistic
corrections) by $V$, our coupled channel Hamiltonian takes the
form:
\be
H \approx \frac{p^2}{2m} + V + H_\gamma -\frac{e}{m}p\cdot A(0) +\dots
\ee
where the dots represent any new instantaneous potentials generated
by interaction with the transverse photon sector. 

The coupled channel problem can be reduced to an effective 
energy-dependent Hamiltonian acting only on the electron subspace.  In
general, if the coupled Schr\"{o}dinger equation has the form:
\be
\left(
\begin{array}{cc}
H_e & H^\prime \\
H^{\prime\dagger} & H_e + H_{\gamma} \\
\end{array}
\right)
\left(
\begin{array}{c}
\psi_e \\
\psi_{e\gamma}
\end{array}
\right) 
=E
\left(
\begin{array}{c}
\psi_e \\
\psi_{e\gamma}
\end{array}
\right),
\ee
then the effective Hamiltonian is
\be
\label{eq: Hcoupled}
H = H_e + H^\prime(E-H_e - H_{\gamma})^{-1}H^{\prime\dagger} .
\ee 
In the present case, 
\bea
H_e &=& \frac{p^2}{2m} + V \nl
H^\prime &=& -\frac{e}{m}{\bf p}\cdot {\bf A}(0),
\eea
and $H_\gamma$ is given by Eq.(\ref{eq: Hg}). 
Eq.(\ref{eq: Hcoupled}) can be evaluated explicity.  Including
radiative photon modes of momentum~\footnote{
Taking $\Lambda_\gamma\approx m$ instead of $\Lambda_\gamma\approx m\alpha$
avoids the appearance of factors $\ln{\alpha}$ in higher order contact terms, 
but is not essential.
The difference will be accounted for by the remaining instantaneous 
interactions---see Eq.(\ref{eq: VF1}).
} $q\le \Lambda_\gamma\approx m$, and 
assigning the photon a mass $\lambda$ to control infrared 
divergences, 
\bea
\label{eq: Lamb}
H &=& H_e + \frac{\alpha}{\pi m^2}p^i\left[
	-\int_0^{\Lambda_\gamma}dq\frac{q^2}{q_\lambda^2}
		\left(\delta^{ij}-\frac{q^iq^j}{q_\lambda^2}\right)\right.\nl
&&\left.	
	+\int_0^{\Lambda_\gamma}dq\frac{q^2}{q_\lambda^2}
		\left(\delta^{ij}-\frac{q^iq^j}{q_\lambda^2}\right)
		\left(\frac{q_\lambda}{E-H_e-q_\lambda}+1\right)\right]p^j,
\eea
with $q_\lambda\equiv \sqrt{q^2+\lambda^2}$.  

The final step in our
analysis is to compute any remaining instantaneous potentials, by
comparing our effective theory to QED.  We do this by
examining the $\order(\alpha)$ correction to the QED scattering
amplitude, which can be written in terms of vertex form factors, and
a vacuum polarization contribution:
\be
\label{eq: Ttrue}
T_{\rm true} = \frac{-4\pi(Z\alpha)}{q^2}F_1
	+\frac{4\pi(Z\alpha)}{4m^2}
		\left(1-\frac{2i\,{\bf l}\times {\bf k}\cdot{\bf \sigma}}{q^2} \right)F_2
	+\frac{-4\pi(Z\alpha)}{q^2}
		\left(\frac{\alpha}{15\pi}\frac{q^2}{m^2}\right) ,
\ee
with
\bea
F_1 &=& 1 - \frac{\alpha}{3\pi m^2}{q^2}\left(\ln\frac{m}{\lambda}-\frac{3}{8}
	+\order(\lambda) \right) +\order(q^4) \\
F_2 &=& \frac{\alpha}{2\pi} + \order(q^2)
\eea
To this we must compare the $\order(\alpha)$ contribution coming
from Eq.(\ref{eq: Lamb})
\footnote{
Here wavefunction renormalization must be taken into account, since
the potential is energy-dependent. 
}:
\be
\label{eq: Teff}
T_{\rm eff} = \frac{-4\pi(Z\alpha)}{q^2}\left[\frac{-\alpha}{3\pi m^2}q^2
	\left(\ln\frac{2\Lambda_\gamma}{\lambda}-\frac{5}{6}\right)\right].
\ee
Comparing Eqs.(\ref{eq: Ttrue}) and (\ref{eq: Teff}), we see that 
potentials:
\bea
\label{eq: VF1}
V_{F1} &=& \frac{4\alpha(Z\alpha)}{3m^2}
		\left(\ln\frac{m}{2\Lambda_\gamma}+\frac{11}{24}\right)
		\delta_\Lambda^3(r) \\
V_{F2} &=& \frac{(Z\alpha)}{4m^2}\frac{\alpha}{2\pi}
		\left(4\pi\delta_\Lambda^3(r)-2\frac{1}{r}
			\left(\frac{1}{r}\right)^\prime 
		{\bf L}\cdot{\bf \sigma}\right) \\
V_{VP} &=& -\frac{4\alpha(Z\alpha)}{15m^2}\delta_\Lambda^3(r),
\eea
associated with $F_1$ and $F_2$ form factors, and vacuum polarization,
respectively, must be added to our effective theory.  Having determined
these potentials, we now set $\lambda=0$, and evaluate the 
$\order(\alpha)$ piece of Eq.(\ref{eq: Lamb})~
\footnote{
The first term in square brackets in Eq.~(\ref{eq: Lamb}) is absorbed
by a mass renormalization.  ``LS'' stands for ``Lamb shift'', since
this potential accounts for the dominant part of this effect.  Note
also that a cutoff in the form ${\rm exp}(-p^2/2\Lambda^2)$
has been included in $V_{LS}$.  
}:
\be
V_{LS}(E) = \frac{2\alpha}{3\pi m^2}e^{-\frac{p^2}{2\Lambda^2}}\,p^i\,(\frac{p^2}{2m}+V_C-E)
	\ln\frac{\Lambda_\gamma}{\frac{p^2}{2m}+V_C-E}\,p^ie^{-\frac{p^2}{2\Lambda^2}}.
\ee

We can now present the complete effective Hamiltonian, including
leading relativistic and radiative corrections:
\be
H_{\rm eff} = \frac{p^2}{2m} + V_C + V_{\rm rel} + V_{\rm rad} + V_{\rm ct},
\ee
where $V_C$ is the cutoff Coulomb potential, 
$V_{\rm rel} = V_K(E) + V_D + V_{SO}$
incorporates relativistic corrections,
$V_{\rm rad} = V_{LS}(E) + V_{F1} + V_{F2} + V_{VP}$
contains the radiative corrections, and
$V_{\rm ct} = (d_1^{(1)}/m^2 + (Z\alpha)d_1^{(2)}/m^2)
	\delta^3_\Lambda(r)$
is the counterterm potential.  For this application, we have
(Eqs. (\ref{eq: d11}),(\ref{eq: d12rel}) ) 
\bea
\nonumber
d_1^{(1)} &=& -2\pi\frac{m^2}{\Lambda^2},  \nl
d_1^{(2)} &=& \sqrt{\pi}\left(-\frac{10}{3}\left(\frac{m}{\Lambda}\right)^3
-\frac{m}{\Lambda} +\frac{1}{8}\frac{\Lambda}{m}\right).
\eea

The spectrum of this Hamiltonian can be readily determined using 
matrix diagonalization  
\footnote{
The necessary 
matrix elements of the various operators
between gaussian basis functions are listed in the appendix.
}, and reproduces (Table \ref{table: lamb}) 
the well-known result through $\order(m\alpha^5)$~
\cite{Sapirstein}~:
\bea
\label{eq: lamb}
E(n,j,l) &=& -\frac{m(Z\alpha)^2}{2n^2}
	-\frac{m(Z\alpha)^4}{2n^3}\left(\frac{1}{j+1/2}-\frac{3}{4n}\right) \nl
&&+ \frac{m\alpha(Z\alpha)^4}{\pi n^3}
	\left[\delta_{l,0}\left(\frac{8}{3}\ln{(Z\alpha)^{-1}}
			+ \frac{10}{9}-\frac{4}{15} \right)\right.\nl	
&&\left.\qquad
	+(1-\delta_{l,0})\left(\frac{j(j+1)-l(l+1)-3/4}{2l(l+1)(2l+1)}
					\right)
		-\frac{4}{3}\ln k_0(n,l)
	\right].
\eea

\begin{table}
\begin{center}
\begin{tabular}{lcc}
\hline
$\alpha$ & $(E(1,1/2,0)-E^\prime(1))/(-4\alpha^5/3\pi n^3)$ & 
$(E(2,1/2,0)-E^\prime(2))/(-4\alpha^5/3\pi n^3)$ \\
\hline
$0.04$ & $2.79656$ & $2.51928$ \\
$0.02$ & $2.88486$ & $2.65806$ \\
$0.01$ & $2.92986$ & $2.72998$ \\
\hline
\end{tabular}
\caption{
$S$~-state energy levels with relativistic and radiative 
corrections, at $\Lambda=\Lambda_\gamma=m$.  
Here 
$E^\prime(n)/m=$ $-(Z\alpha^2)/2n^2-(Z\alpha)^4/2n^3(1-3/4n)
+\alpha(Z\alpha)^4/\pi n^3(8/3\ln\alpha^{-1}+10/9-4/15)$
is the part of Eq.(\ref{eq: lamb}) excluding $\ln k_0$.  
At $\alpha \to 0$ the second and third columns converge to 
$\ln k_0(1,0)=2.98413$ and $\ln k_0(2,0)=2.81177$, respectively.
}
\label{table: lamb}
\end{center}
\end{table}
It is worth noting that higher order terms , including ones
containing $\ln(\alpha)$, appear automatically 
when the theory is solved nonperturbatively.   This
can be important in high orders when the appearance of 
factors $(\ln\alpha)^n$,
for large enough $n$, causes poor convergence of series expansions.  

\,

\noindent{\bf Acknowledgements}. The work presented here was done in 
collaboration with Peter Lepage.  I also thank 
Jonathan Sapirstein for useful conversations, and Patrick Labelle 
for a proofreading of the manuscript.

\begin{appendix}
\section*{Appendix: Matrix Elements for Gaussian Basis}
Following is a list of matrix elements between basis functions
\be
\Phi^i_{lm}({\bf r}) = Y_{lm}(\theta,\phi) r^l e^{-\frac{r^2}{2R_i^2}}.
\ee
It is convenient to define the quantity
$R_{ij} = (R_i^{-2}+R_j^{-2})^{-1/2}$.
Then the matrix elements are: (a factor $\delta_{m_i,m_j}\delta_{l_i,l_j}$ 
is suppressed)
\bea
\left[r^n\right]_{ij} 
&=& 	2^{l+(n+1)/2}\Gamma\left(\frac{3}{2}+l+\frac{n}{2}\right)
		R_{ij}^{3+2l+n} \\
\left[p^2\right]_{ij}
&=& 	\frac{l+3/2}{R_{ij}^2}\left[r^0\right]_{ij} 
		-\frac{1}{2}\left(\frac{1}{R_i^4}+\frac{1}{R_j^4}\right)
		\left[r^2\right]_{ij} \\
\left[\left(\frac{1}{r}\right)_\Lambda\right]_{ij}
&=&	\frac{1}{\Lambda^{2(l+1)}}2^{l}\left(-\frac{d}{da}\right)^l
		\left[\frac{1}{a\sqrt{1+a}}\right]_{a=1/(\Lambda^2 R_{ij}^2)}
			\\
\left[\delta^3_\Lambda({\bf r})\right]_{ij}
&=&
\frac{2^{l+1/2}}{4\pi\Lambda^{2l}}\sqrt{\frac{2}{\pi}}
	\Gamma\left(\frac{3}{2}+l\right)
	\frac{1}{\left(1+\frac{1}{\Lambda^2R_{ij}^2}\right)^{l+3/2}} \\
\left[-\nabla^2\delta^3_\Lambda({\bf r})\right]_{ij}
&=&
3\Lambda^2\left[\delta^3_\Lambda({\bf r})\right]_{ij}
-\frac{2^{l+3/2}}{4\pi\Lambda^{2l-2}}\sqrt{\frac{2}{\pi}}
	\Gamma\left(\frac{5}{2}+l\right)
	\frac{1}{\left(1+\frac{1}{\Lambda^2R_{ij}^2}\right)^{l+5/2}} \\
\left[\frac{1}{r}\left(\frac{1}{r}\right)_\Lambda\right]_{ij}
&=& 
\frac{1}{\Lambda^{2l}}\left\{
	-2^{l-1}\left(-\frac{d}{da}\right)^l
		\left[\ln\frac{\sqrt{1+a}+1}{\sqrt{1+a}-1}\right]
			_{a=1/(\Lambda^2 R_{ij}^2)} \right.\nl
&&\left. +\frac{2^l}{\sqrt{\pi}}\Gamma\left(l+\frac{1}{2}\right)
		\frac{1}{\left(1+\frac{1}{\Lambda^2 R_{ij}^2}\right)^{l+1/2}}
	\right\} 
			\\
\eea

For the operator $p^i$ appearing in $V_{LS}$, we concentrate
on the ground state, with $l=0$.  To first order in $V_{LS}$,
we need only consider coupling between $S$- and $P$- states 
($S \to P\to S$ transitions)\footnote{
Higher orders will introduce $S\to P\to D \to P\to S$ 
transitions, but these are suppressed by several powers of $\alpha$.
}.  Here the sum over spherical harmonics results in the replacement
$\sum_i p^i(\cdots)p^i \to -(ip_r)(\cdots)(ip_r)$, where $ip_r=d/dr$,
and:
\be
\left[ip_r\right]_{ij}
=\delta_{l_i,1}\delta_{l_j,0}\left(-\frac{[r^1]_{ij}}{R_j^2}\right)
	+\delta_{l_i,0}\delta_{l_j,1}\left(\frac{[r^1]_{ji}}{R_i^2}\right).
\ee
Radially excited states ($l>0$) can be treated similarly; for instance
$l=1$ requires a sum over couplings between $P$- and $S$- and between
$P$- and $D$-states.   


The $R_i$ should take values which cover the range from $\sim 1/\Lambda$
to $\sim 1/(m\alpha)$.  For example, 
the values of Table~\ref{table: coulomb}
were generated using $R_i=R_{\rm min}(R_{\rm max}/R_{\rm min})^{i/N}$,
with $i=0..N$, $N=60$, $R_{\rm min}=0.1/\Lambda$, $R_{\rm max}=100/(m\alpha)$
and $\Lambda=m$, $\alpha=0.02$.  
For problems with more than one angular momentum channel, a 
separate set of 
basis functions should be used for both channels, e.g. $N_S$ S-states
and $N_P$ P-states for the hydrogen atom problem with radiative 
corrections.

\end{appendix}


\begin{thebibliography}{99}

\bibitem{Renormalizing} G.P. Lepage,
``How to Renormalize the Schroedinger Equation,''
in: Nuclear Physics, Proceedings of the VIII Jorge Andr\'{e} Swieca 
Summer School, January, 1995, edited by C.A. Bertulani et. al.
(World Scientific, Singapore, 1997), p. 135. [nucl-th/9706029]

\bibitem{ops} R. Hill, G.P. Lepage, to be published. [hep-ph/0003277]

\bibitem{NRQED} NRQED was introduced in 
W.E. Caswell, G.P. Lepage, Phys. Lett. {\bf 167B}, 437 (1986). 
A more detailed description can be found in T. Kinoshita, M. Nio,
Phys.Rev. {\bf D53} 4909 (1996). 

\bibitem{NRecipes} W.H. Press, B.P. Flannery, S.A. Teukolsky and
W.T. Vetterling, {\it Numerical Recipes in C} 
(Cambridge University Press, Cambridge, 1988). 

\bibitem{Bethe} H.A. Bethe and E.E. Salpeter, {\it Quantum Mechanics
of One- and Two-Electron Atoms} (Springer-Verlag, Berlin, 1957).  

\bibitem{Sapirstein}
J. Sapirstein and D.R. Yennie in: {\it Quantum Electrodynamics}, 
edited by T. Kinoshita (World Scientific, Singapore, 1990), p. 579.

\end{thebibliography}
\end{document}